\PassOptionsToPackage{table}{xcolor}
\documentclass[a4paper,fleqn]{cas-dc}

\usepackage{amsmath,amssymb,amsfonts}
\usepackage{algorithmic}
\usepackage{array}
\usepackage{textcomp}
\usepackage{stfloats}
\usepackage{url}
\usepackage{graphicx}
\usepackage{pgfplots}
\pgfplotsset{compat=1.18}

\usepackage{multirow}
\usepackage{booktabs}
\usepackage{colortbl}
\usepackage{tabularx}
\usepackage{makecell}
\usepackage{multicol}

\usepackage{hyperref}
\usepackage{tcolorbox}
\usepackage{fancybox}
\usepackage{framed}
\usepackage{placeins}

\usepackage{fontawesome}
\usepackage{orcidlink}
\usepackage{enumitem}
\usepackage{csquotes}
\usepackage{soul}
\usepackage{comment}

\usepackage{changepage}

\usepackage[table]{xcolor}
\definecolor{methodFocus}{HTML}{0173B2}
\definecolor{methodInterview}{HTML}{DE8F05}
\definecolor{methodSurvey}{HTML}{029E73}

\definecolor{tgreen}{HTML}{c6ecd1}
\definecolor{tblue}{HTML}{bdd7ee}
\definecolor{tgray}{HTML}{e7e6e6}

\tcbuselibrary{theorems}

\tcbset{
    colback=gray!5,
    colframe=gray!40,
    colbacktitle=gray!25,
    coltitle=black,
    boxrule=0.4pt,
    arc=2pt,
    left=4pt,
    right=4pt,
    top=4pt,
    bottom=4pt,
}

\newtcolorbox{userreview}[1][]{
    colback=gray!5,
    colframe=gray!40,
    fonttitle=\bfseries,
    title=#1
}

\newtcbtheorem{customexample}{\faAlignJustify{} \footnotesize Example}
{
    colback=blue!5,
    colframe=blue!35!black,
    fonttitle=\footnotesize \bfseries,
    fontupper=\footnotesize
}{th}

\newenvironment{summary}{%
\vspace{4pt}\par
\noindent\begin{minipage}[tl]{\linewidth}
	\MakeFramed{\advance\hsize-\width\FrameRestore}%
	\noindent\hspace{-4.55pt}% disable indenting first paragraph
	\begin{adjustwidth}{}{4pt}%
	}
	{%
	\end{adjustwidth}\endMakeFramed%
\end{minipage}
}

\newcounter{findingctr}
\newcommand{\finding}[1]{%
  \refstepcounter{findingctr}%
  \begin{summary}%
    \small\textbf{Finding \thefindingctr:} #1%
  \end{summary}%
}

\graphicspath{{./figures/}}

\ExplSyntaxOn\bool_gset_true:N \g_stm_nologo_bool\ExplSyntaxOff
\ExplSyntaxOn
\cs_set:Npn \__first_footerline: {
  \group_begin:
  \small
  \sffamily
  \ifnum\theblind>0\relax
  \else 
	\__short_authors: :~
  \fi
  { \rmfamily \itshape Preprint~ submitted ~to ~Information~and~Software~Technology }
  \group_end:
}
\ExplSyntaxOff
\let\ISTmathasterisk\ast
\renewcommand{\ast}{\text{\normalfont *}}
\begin{document}

\shorttitle{Developing a Multidimensional Taxonomy of Explanation Needs}
\shortauthors{Obaidi et al.}

\title[mode=title]{Developing and Refining a Multidimensional Taxonomy of Explanation Needs in App Reviews}

\author[1]{Martin Obaidi}[orcid=0000-0001-9217-3934]
\ead{martin.obaidi@inf.uni-hannover.de}
%\credit{Conceptualization, Data curation, Formal analysis, Investigation, Methodology, Resources, Software, Validation, Visualization, Writing -- original draft, Writing -- review \& editing}
\cormark[1]

\author[1]{Marc Herrmann}[orcid=0000-0002-3951-3300]
\ead{marc.herrmann@inf.uni-hannover.de}
%\credit{Writing -- review \& editing}

\author[1]{Kurt Schneider}[orcid=0000-0002-7456-8323]
\ead{kurt.schneider@inf.uni-hannover.de}
%\credit{Funding acquisition, Supervision}

\cortext[cor1]{Corresponding author: Martin Obaidi, martin.obaidi@inf.uni-hannover.de}

\address[1]{
Leibniz Universität Hannover,
Software Engineering Group,
Hannover, Germany
}

\begin{comment}
\shortauthors{Anonymous Authors}
\author[1]{Anonymous Authors}
\address[1]{Anonymous Institution}
\end{comment}

\begin{abstract}
\mbox{}\par
\noindent\textit{Context.}
A single app review can express an explanation need that concerns an update, a feature change, and a particular system component.

\noindent\textit{Objectives.}
We develop a taxonomy that captures these aspects together for each explanation need.

\noindent\textit{Methods.}
Two researchers combined existing classifications with joint coding of a corpus of 4{,}495 reviews, including 2{,}185 with a recorded need. An expert pre-study and independent coding of 104 purposively selected reviews informed refinement. We also analyzed three raters' assignments to aligned needs in a separate dataset of 5{,}004 reviews.

\noindent\textit{Results.}
The taxonomy distinguishes \emph{Time Aspect}, \emph{Unexpected System Behavior \& Bugs}, \emph{Software Feature}, and \emph{System Aspect}. In the formative evaluation, the two raters agreed on 314 of 416 dimension assignments (75.48\%) and on all four labels for 35 of 104 reviews (33.65\%). Cohen's kappa ranged from 0.467 to 0.911 across dimensions. For the refined taxonomy in the separate dataset, category-level Fleiss' kappa ranged from 0.41 to 0.71 for the first three dimensions and from 0.25 to 0.83 for System Aspect.

\noindent\textit{Conclusion.}
The four labels let analysts group needs by release context, behavior, feature, or subject. Agreement varied by category. The evaluations leave open how refinement changed reliability and whether classification improves engineering outcomes.
\end{abstract}


\makeatletter
\def\Hy@Warning#1{}
\makeatother

\maketitle
\let\ast\ISTmathasterisk

\section{Introduction}
\label{sec:introduction}

Post-release user feedback is an important source of information for software engineering. App reviews reveal bug reports, feature requests, usage problems, and user experiences that emerge after interaction with deployed software~\cite{Pagano2013,maalej2016,guzman2014appreviewsfeature,iacob2013appreviews}. Such feedback can inform requirements decisions, maintenance, documentation, and product planning. Its practical value, however, depends on whether development teams can transform heterogeneous review text into information that supports analysis and action.

One relevant type of information in app reviews is an \emph{explanation need}, a knowledge gap concerning an aspect of a software system that a user seeks to understand. Explanation needs may be expressed explicitly through questions or requests for information, or implicitly through confusion, surprise, uncertainty, or descriptions of behavior that the user cannot explain~\cite{unterbusch2023explanation,droste2024taxonomy}. They can concern feature changes, unexpected behavior, permissions, intended functionality, or organizational decisions.

Detecting an explanation need alone does not capture enough information for subsequent software engineering activities. A feature-related need may inform requirements or documentation, unexpected behavior may require technical investigation, and security or support concerns may involve different stakeholders. A useful representation should therefore preserve several properties of the same explanation need.

Existing work provides important foundations. App-review mining distinguishes categories such as bug reports, feature requests, user experiences, and information-seeking statements~\cite{maalej2016,S.Panichella.2015}. Explainability research has shown that explanation needs occur in app reviews and can be detected automatically~\cite{unterbusch2023explanation}. Existing taxonomies further describe recurring subjects such as system behavior, interaction, domain knowledge, privacy and security, and user interface concerns~\cite{droste2024taxonomy,droste2025operationaltaxonomy}. These approaches, however, primarily organize a need from one classification perspective.

This becomes limiting when one explanation need contains several complementary properties. Consider the review ``\emph{Why was the offline mode removed after the latest update?}'' It contains a temporal reference, concerns a feature modification, and refers to a system aspect. Assigning the need to a single category loses information, while combining all properties in one hierarchy creates overlap.

We therefore develop and refine a multidimensional taxonomy for explanation needs in app reviews. The taxonomy represents each identified need along four complementary dimensions: \emph{Time Aspect}, \emph{Unexpected System Behavior \& Bugs}, \emph{Software Feature}, and \emph{System Aspect}. This structure preserves different properties of the same need while remaining independent of platform-specific metadata and organization-specific prioritization criteria.

The taxonomy combines concepts from existing classifications with a joint bottom-up analysis of app reviews from an existing gold-standard dataset~\cite{goldstandard-explain-zenodo2024}. We formatively evaluated the initial taxonomy through an expert pre-study and an independent inter-rater application involving 104 reviews. The resulting agreement values and disagreement patterns informed the refinement of categories and annotation guidelines. Complementary post-refinement validation evidence is drawn from a separate public multidimensional dataset of 5{,}004 reviews~\cite{obaidi2026multigolddataset}. Its construction and broader analysis are separate research outputs. An industrial elicitation study used an extended coding scheme~\cite{obaidi2025elicitation,obaidi2026eliciting}.

This article makes the following contributions:

\begin{itemize}
    \item A content-based taxonomy with four dimensions, category definitions, annotation rules, and mappings to two existing classifications.

    \item A documented derivation from a criteria catalog and a formative evaluation linking disagreements to changes in the taxonomy.

\end{itemize}

The remainder of this article reviews the relevant foundations, presents the study design, describes the development and refinement of the taxonomy, and discusses validation evidence, implications, limitations, and future work.

\section{Background and Related Work}
\label{sec:background}

This section introduces explanation needs and multidimensional classification, reviews related taxonomies and app-review research, and positions the present study within these areas.

\subsection{Explainability and Explanation Needs}
\label{sec:background:explainability}

Explainability refers to the ability of a software system to be explained~\cite{chazette2020explainability}. It concerns the relationship between a system, the aspect to be explained, the addressee, the context, the explainer, and the information provided~\cite{chazette2021exploring}. Although explainability research is often associated with artificial intelligence and model predictions~\cite{ribeiro2016should,Adadi2018,das2020opportunities}, explanation needs also occur for conventional software systems.

What users seek to understand depends on their knowledge, context, and mental model of the system~\cite{kohl2019explainability,ramos2021modeling,anders2022mentalmodels}. Preferences regarding explanation detail and format can likewise vary between users and situations~\cite{obaidi2025appknowledge,obaidi2025mood}. Providing explanations without considering these needs can increase cognitive load or impair usability~\cite{chazette2020explainability,chazette2021exploring}. Related work therefore also addresses the quality and measurement of implemented explanations~\cite{deters2024qualitymodel,deters2025metrics}.

Following Unterbusch et al.~\cite{unterbusch2023explanation}, we use the term \emph{explanation need} for a knowledge gap expressed in user feedback. Explanation needs can be explicit or implicit~\cite{droste2024taxonomy}. An explicit need may be expressed as a direct question such as ``\emph{Why does the app require access to my location?}'' An implicit need may instead appear through confusion or surprise, for example ``\emph{The app asks for my location every time I open it, and I do not understand what it is used for.}'' Throughout this article, we use \emph{explanation need} and treat the term \emph{explainability need} used in related work as referring to the same concept.

\subsection{Taxonomies and Classification Perspectives}
\label{sec:background:taxonomies}

A taxonomy organizes entities according to shared characteristics and supports common terminology, comparison, and the identification of gaps within a domain~\cite{usman2017taxonomies,vegas2009maturing}. In requirements engineering, taxonomies can support stakeholder discussions, provide terminology for formulating requirements, and structure information extracted from user feedback~\cite{lauenroth2017words,droste2024taxonomy,unterbusch2023explanation}.

Many software engineering taxonomies use hierarchical structures in which general categories are divided into more specific subcategories~\cite{usman2017taxonomies}. Such structures work well when categories represent one classification perspective. They become less suitable when an entity has several complementary properties. An explanation need can simultaneously concern a recent update, a changed feature, and the user interface.

A multidimensional taxonomy separates these properties into distinct classification perspectives. Each dimension represents one property of the classified entity and contains its own categories. The same explanation need can therefore receive one label in each dimension without forcing complementary properties into competing categories.

\subsection{Existing Taxonomies Related to Explainability}
\label{sec:background:classifying}

Taxonomies related to explainability differ in their objects of classification. Some classify explainability methods, others classify explanations provided by systems, and a smaller group focuses on users' explanation needs.

Speith~\cite{speith2022XAITaxonomies} reviewed taxonomies of explainable artificial intelligence methods and provided recommendations for their development and use. These taxonomies primarily distinguish methods for explaining models or decisions rather than the aspects of software systems that users seek to understand.

Tsakalakis et al.~\cite{tsakalakis2022taxonomy} propose a multidimensional taxonomy for explanations provided by applications. Their dimensions include the source, trigger, content, scope, and intended recipient of an explanation. Although their object of classification is the provided explanation rather than the explanation need, their work demonstrates the use of complementary classification perspectives for explainability-related artifacts.

Droste et al.~\cite{droste2024taxonomy,droste2025operationaltaxonomy} developed a taxonomy of explanation needs in everyday software systems based on an online survey with 84 participants and 315 explicit and implicit explanation needs. Its top-level categories are \emph{System Behavior}, \emph{Interaction}, \emph{Domain Knowledge}, \emph{Privacy and Security}, and \emph{User Interface}. The taxonomy provides terminology for recurring subjects that users seek to understand and shows that explanation needs extend beyond artificial intelligence.

Unterbusch et al.~\cite{unterbusch2023explanation} focus specifically on explanation needs in app reviews. They manually analyzed 1{,}730 reviews from eight applications and distinguish between \emph{Primary Concerns} and \emph{Secondary Concerns}. Primary concerns include \emph{Training}, \emph{Interaction}, and \emph{Business}, while secondary concerns include \emph{Dissatisfaction} and \emph{Errata}. Their study also demonstrates that explanation needs in app reviews can be detected automatically.

\subsection{Mining and Classifying App Reviews}
\label{sec:background:appreviews}

App stores provide a continuous channel through which users communicate experiences, problems, and requirements after interacting with deployed software~\cite{Pagano2013}. Pagano and Maalej~\cite{Pagano2013} analyzed more than one million App Store reviews and identified feedback related to ratings, user experience, requirements, and community interaction.

Subsequent work has developed techniques for structuring this feedback. Maalej et al.~\cite{maalej2016} classify reviews into bug reports, feature requests, user experiences, and ratings. Panichella et al.~\cite{S.Panichella.2015} distinguish sentences that provide information, seek information, request features, or report problems. Other approaches identify informative reviews~\cite{AR-Miner}, extract discussed features and sentiment~\cite{guzman2014appreviewsfeature}, and cluster related feedback~\cite{villarroel2016mobileappreviews}. User feedback is selective~\cite{anders2023userfeedback}. Metadata and linguistic indicators can complement the review text, but the reliability of sentiment estimates depends on the dataset, platform, and analysis tool~\cite{obaidi2025trustworthy,obaidi2022crossplatform}.

Unterbusch et al.~\cite{unterbusch2023explanation} extend app-review mining to the identification of explanation needs. Obaidi et al.~\cite{obaidi2025explanationneeds} subsequently use explanation-need categories in an industrial setting to assign detected needs to the internal teams responsible for answering them.

\subsection{Synthesis and Positioning of This Work}
\label{sec:background:synthesis}

Prior work establishes explanation needs as relevant knowledge gaps, provides taxonomies for recurring subjects and concerns, and demonstrates that app reviews can be mined for software engineering information. A gap remains in representing several complementary properties of the same identified explanation need simultaneously. Existing explanation-need taxonomies primarily organize needs from one classification perspective, while app-review classifications generally characterize the purpose or content of feedback. Our work addresses this gap by treating the individual explanation need as the unit of classification and representing it through four complementary dimensions. Table~\ref{tab:positioning} summarizes the relation to the two closest explanation-need classifications.

\begin{table*}[t]
\centering
\caption{Positioning of the present work relative to the closest explanation-need classifications.}
\label{tab:positioning}
\begin{tabularx}{\textwidth}{p{0.17\textwidth} p{0.22\textwidth} p{0.24\textwidth} X}
\toprule
\textbf{Approach} & \textbf{Empirical basis} & \textbf{Primary representation} & \textbf{Main distinction} \\
\midrule
Droste et al.~\cite{droste2024taxonomy,droste2025operationaltaxonomy}
& 84 participants and 315 explanation needs
& Recurring subjects of explanation needs
& Focuses on what users seek to understand. \\

Unterbusch et al.~\cite{unterbusch2023explanation}
& 1{,}730 app reviews from eight applications
& Primary and secondary concerns
& Focuses on concerns expressed in app reviews. \\

This work
& 4{,}495 development reviews in v20, including 2{,}185 with recorded needs
& Four complementary dimensions per explanation need
& Represents several properties of the same need simultaneously. \\
\bottomrule
\end{tabularx}
\end{table*}

The resulting taxonomy complements existing detection and classification approaches by preserving temporal, behavioral, feature-related, and system-related properties for subsequent software engineering analysis.

For example, Droste's \emph{Unexpected Behavior} indicates a discrepancy. Identifying data loss, a wrong result, or a performance change requires reading the need. Unterbusch's \emph{Design Decision} can identify the subject as \emph{Development Rationale}, whereas its training categories for instruction, available features, and feature effects can all map to \emph{Feature Confusion}. Conversely, the latter label alone cannot recover which training concern was recorded. Unterbusch also permits primary and secondary concerns. Section~\ref{sec:taxonomy:final} illustrates what the archived labels capture on the same cases.

\section{Study Design}
\label{sec:studydesign}

This section presents the research questions and describes the procedure, data sources, and analysis methods used to develop and formatively evaluate the taxonomy.

\subsection{Research Goal and Research Questions}
\label{sec:studydesign:goal}

We formulate our research goal using the goal definition template by Wohlin et al.~\cite{wohlin2012experimentation} based on Basili and Rombach~\cite{basili1988goaltemplate}:

\begin{quote}
\textbf{Research Goal:} \emph{Analyze} criteria and categories for characterizing explanation needs \emph{for the purpose of} developing and refining a multidimensional taxonomy that supports the structured analysis of post-release user feedback \emph{with respect to} coverage, conceptual clarity, and consistent applicability \emph{from the point of view of} software engineering researchers and practitioners \emph{in the context of} explanation needs expressed in app reviews.
\end{quote}

\noindent Coverage and consistent applicability are design goals. Our questions examine how the taxonomy was developed and where its use produced disagreement. The evaluation does not estimate population coverage.

Based on this goal, we investigate the following research questions:

\begin{itemize}
    \item[\textbf{RQ1}] Which criteria can be derived from selected prior classifications, related work, and app reviews for characterizing explanation needs?

    \item[\textbf{RQ2}] How can suitable content-level criteria be organized into a multidimensional taxonomy of explanation needs?

    \item[\textbf{RQ3}] Which ambiguities emerge when the initial taxonomy is reviewed and independently applied, and how can the taxonomy be refined accordingly?
\end{itemize}

RQ1 addresses the broader design space of potential criteria. RQ2 concerns the selection and organization of suitable content-level criteria. RQ3 examines ambiguities in the initial taxonomy and how the resulting evidence informs its refinement.

\subsection{Study Procedure}
\label{sec:studydesign:procedure}

\begin{figure*}[t]
    \centering
    \includegraphics[width=\textwidth]{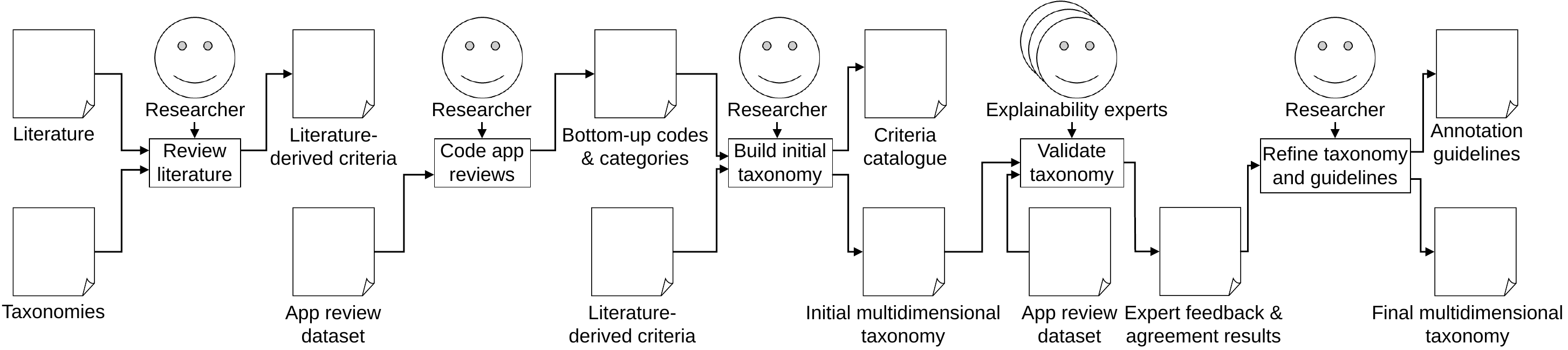}
    \caption{FLOW model of the taxonomy-development procedure based on the notation by Stapel et al.~\cite{stapel2009flow}. Reused agreement evidence from the multidimensional dataset and complementary application-based evidence from the industrial study (Sections~\ref{sec:taxonomy:postrefinement-results} and~\ref{sec:taxonomy:application}) concern the final artifact and are therefore not shown in the development model.}
    \label{fig:study-procedure}
\end{figure*}

Figure~\ref{fig:study-procedure} provides an overview of the taxonomy-development procedure. We first identified candidate criteria from existing classifications and related literature and complemented them through a joint bottom-up analysis of app reviews from an existing gold-standard dataset~\cite{goldstandard-explain-zenodo2024}. The resulting criteria were consolidated into an initial multidimensional taxonomy.

We formatively evaluated the initial taxonomy through an expert pre-study and an independent inter-rater application. Expert feedback, agreement results, and disagreement patterns informed the subsequent refinement and the development of annotation guidelines.

The procedure follows established guidance for taxonomy development in information systems and software engineering~\cite{nickerson2013taxonomy,usman2017taxonomies}. It combines conceptual-to-empirical and empirical-to-conceptual development and uses the content-level characterization of individual explanation needs as its meta-characteristic.

\subsection{Gold-Standard App Review Dataset}
\label{sec:studydesign:dataset}

The empirical basis for the bottom-up analysis was an app-review dataset originally sampled to a target of 4{,}500 reviews and labeled for explanation needs~\cite{goldstandard-explain-zenodo2024}. The dataset was constructed from an initial collection of 90{,}000 reviews using rule-based sampling intended to cover explicit, implicit, and absent explanation needs.

To count needs in the development corpus, we used its consolidated version 20 (worksheet \texttt{dataset\_with\_codes}, 26 October 2026). It contains 4{,}495 review records with unique review keys. The final human labels identify 2{,}185 reviews with a need (1{,}457 explicit and 728 implicit) and 2{,}310 without one. The historical sampling field and the final human label serve different purposes. Each entry in the ordered category list represents one recorded need, including repeated entries of the same category. For every review we checked the list against the eight separate code fields and the recorded total. Across all reviews we counted 3{,}298 needs. Exactly 1{,}399 reviews had one need and 786 had more than one: 574 had two, 137 had three, 46 had four, 20 had five, eight had six, none had seven, and one had eight. Multiple needs therefore occurred in 17.49\% of all 4{,}495 reviews and 35.97\% of the 2{,}185 positive reviews. These are counts of the recorded needs in the consolidated corpus.

\subsection{Identification of Candidate Criteria}
\label{sec:studydesign:criteria}

We used two complementary approaches to identify candidate criteria. The first drew on existing classifications and related literature. The second used bottom-up coding of explanation needs in the app-review dataset.

\subsubsection{Literature-Derived Criteria}

We examined existing taxonomies of explanation needs and related classifications of app reviews to identify concepts that could characterize the content or context of an explanation need. This was a targeted analysis of work closely related to our object of study rather than a systematic literature review.

The reviewed literature contributed concepts such as system behavior, interaction, domain knowledge, privacy and security, user interface, training, business concerns, and prioritization information~\cite{droste2024taxonomy,droste2025operationaltaxonomy,unterbusch2023explanation}. Research on app-review analysis additionally contributed metadata and text-derived criteria such as app version, review length, sentiment, linguistic indicators, and review statistics. Prior evidence of only weak associations between app metadata and explanation needs further supported treating such information as supplementary rather than as content-level dimensions~\cite{obaidi2025appfeatures}.

These concepts were collected as candidate criteria without assuming that each would become part of the taxonomy.

\subsubsection{Bottom-Up Coding of App Reviews}

The bottom-up analysis focused on the explanation need expressed in each relevant review. Two researchers jointly coded passages from the development reviews and iteratively developed and consolidated codes and categories. Unclear cases and category boundaries were discussed until consensus was reached. This was a collaborative qualitative coding process rather than independent double-coding of the complete corpus. Independent applicability was examined later during the formative inter-rater application. We used MAXQDA for the qualitative coding process.\footnote{\url{https://www.maxqda.com}}

The development records include the bottom-up codes, successive code systems, and a MAXQDA project linking codes to passages. Joint coding produced one set of judgments.

Each review received a short code describing the aspect for which the user required an explanation. Similar codes were grouped into categories and iteratively consolidated until further grouping would have combined conceptually different concerns.

The resulting codes covered recurring concerns related to system components, unexpected behavior and bugs, feature changes, temporal references, hardware, commercial aspects, and customer support.

\subsection{Construction of the Initial Taxonomy}
\label{sec:studydesign:construction}

The literature-derived criteria and bottom-up categories were combined into a broader criteria catalog. It included content-level characteristics of explanation needs as well as supplementary information such as app and review metadata, linguistic indicators, inferred properties, mappings to external classifications, and organization-specific information.

Not all criteria were suitable as dimensions of a reusable taxonomy. We therefore assessed whether a criterion described the semantic content of an individual explanation need, could be applied to a single review, could be operationalized categorically by human raters, and remained independent of specific applications and platforms. Section~\ref{sec:taxonomy:criteria} reports the application of these conditions and the resulting selection.

We chose a primary label per dimension so that an analyst could count needs once within that dimension. This sacrifices secondary aspects of a genuinely ambiguous need. Historical materials document no comparative test of a multi-label alternative. The selected content-level criteria formed the dimensions of the initial taxonomy. Each explanation need received one label in every dimension, including a label for cases in which the respective property did not apply. The consistency of these labels and category boundaries was subsequently examined during the formative evaluation.

\subsection{Formative Validation}
\label{sec:studydesign:validation}

The initial taxonomy was formatively evaluated in two steps to identify conceptual weaknesses and problems affecting consistent application. The evaluation was intended to inform refinement rather than to serve as a final evaluation of the resulting taxonomy.

\subsubsection{Expert Pre-Study}

Four explainability experts from the Software Engineering Department at Leibniz University Hannover discussed the interim coding system. Kliti Nikollau presented the objective and hierarchy, then invited the group to propose changes. The discussion informed three changes to the coding system: renaming \emph{No time mentioned} to \emph{Time independent}, grouping bugs and unexpected behavior under a behavior category, and moving time from the root of the hierarchy to its own dimension. Recruitment, individual relationships to the authors, and the recording procedure are undocumented in the available account. The subsequent interviews about criteria prioritization were a separate activity.

\subsubsection{Inter-Rater Application}

Kliti Nikollau and the researcher supervising taxonomy development independently applied the initial multidimensional taxonomy to 104 reviews. The subset was selected to provide broad category coverage. Where possible, at least three reviews were included for each category from the taxonomy by Droste et al.~\cite{droste2024taxonomy,droste2025operationaltaxonomy} and from the initial coding system. The worksheet retains one excerpt and one four-label profile per rater for each of 104 reviews. We linked 103 cases to version 20 by unique source-ID match and one by app and a uniquely matching excerpt. Of the 104 full source reviews, 63 contained one recorded need and 41 contained more than one. Agreement thus describes the retained profiles. The worksheet cannot establish agreement on the segmentation of all needs in the full text.

The prior-taxonomy field contains 23 entries recorded simply as \emph{Business}, 23 as \emph{Unexpected Behavior}, and one as \emph{Consequences}. Some entries combine labels. The target of three examples was a selection aim, and representation remained uneven.

Each rater assigned one label in every taxonomy dimension. Labels indicating non-applicability were treated as regular classification decisions. After the independent application, we compared the assignments for each dimension and qualitatively examined all disagreements to identify recurring causes.

\subsection{Taxonomy Refinement and Annotation Guidelines}
\label{sec:studydesign:refinement}

The final refinement was based on the expert feedback and the quantitative and qualitative results of the inter-rater application. We examined recurring disagreements and unclear category boundaries to identify categories that required consolidation, restructuring, renaming, or clearer definitions.

Based on these findings, we revised the taxonomy and developed annotation guidelines containing definitions, assignment rules, and examples for the final categories. Section~\ref{sec:taxonomy:optimization} reports the concrete refinement decisions and their empirical basis.

\subsection{Complementary Post-Refinement Evidence}
\label{sec:studydesign:postrefinement}

The final taxonomy was subsequently applied by three independent raters to a separately constructed, publicly available gold-standard dataset of 5{,}004 app reviews~\cite{obaidi2026multigolddataset}. The dataset was constructed independently of the development dataset and has no application-level or review-level overlap with it. The three raters were not involved in taxonomy development, and the final category structure and annotation guidelines were fixed before the main annotation.

This article, the development dataset, the separate gold standard, and the industrial study share a first author~\cite{goldstandard-explain-zenodo2024,obaidi2026multigolddataset,obaidi2025elicitation}. Separate here refers to the reviews, apps, and rating exercise. The dataset publication supplies the gold standard and annotation results. This article traces the design and formative refinement of the scheme they used.

We reanalyze category-level agreement from the archived assignments preceding conflict resolution as complementary post-refinement validation evidence. These values characterize the applicability of the final taxonomy in a separate annotation setting. The construction and broader analysis of the dataset are separate research outputs and are not contributions of the present article.

\subsection{Data Analysis}
\label{sec:studydesign:analysis}

For the formative 104-review inter-rater application, we calculated percentage agreement and Cohen's Kappa separately for each taxonomy dimension~\cite{cohen1960agreement}. We additionally report overall percentage agreement across all classification decisions. Kappa values are interpreted descriptively using the benchmarks by Landis and Koch~\cite{landis1977measurement}, shown in Table~\ref{tab:kappa-interpretation}.

We also counted reviews with agreement on all four labels. The 95\% intervals for dimension-level $\kappa$ use the percentile bootstrap from the archived analysis (5{,}000 draws of 104 whole reviews with replacement, seed 42). They describe resampling variability conditional on this purposive set. The Landis--Koch cutoffs are conventions. Interpret differences alongside category prevalence and uncertainty.

\begin{table}[t]
\centering
\caption{Interpretation of Kappa values according to Landis and Koch~\cite{landis1977measurement}.}
\label{tab:kappa-interpretation}
\begin{tabular}{ll}
\toprule
\textbf{$\kappa$} & \textbf{Interpretation} \\
\midrule
$< 0.00$ & Poor \\
$0.00$ to $0.20$ & Slight \\
$0.21$ to $0.40$ & Fair \\
$0.41$ to $0.60$ & Moderate \\
$0.61$ to $0.80$ & Substantial \\
$0.81$ to $1.00$ & Almost perfect \\
\bottomrule
\end{tabular}
\end{table}

For the complementary post-refinement evidence, we recalculate category-level Fleiss' Kappa, Gwet's AC1, and percentage agreement from the three-rater annotation of the separate multidimensional dataset~\cite{fleiss1971measuring,gwet2002kappa,gwet2001handbook,obaidi2026multigolddataset}. The measures are considered jointly because chance-corrected agreement coefficients can be affected by low or highly imbalanced category prevalence.

For the final scheme, we reconstructed the archived binary vote matrices and verified the round estimates against the original analysis exports. For each of 5{,}000 bootstrap replicates (seed 20261124), we sampled the contributing reviews with replacement independently within round, retained all passages and all three votes in each sampled review, calculated Fleiss' $\kappa$ within each round, and averaged the two estimates. We report the 2.5th and 97.5th percentiles. All 12{,}740 passage-by-dimension vote triples matched the original analysis matrices. Comparison with the six raw annotation exports identified 11 passage-by-dimension records with a fourth label entry. The historical matrices retained three votes. Excluding the 11 affected records from their respective dimensions changed each reported mean $\kappa$ by less than 0.003. Positive support counts passages receiving at least one positive vote. Parents and children overlap. These statistics refer to classifying aligned needs, conditional on identification.

Finally, we qualitatively analyzed disagreements from the formative inter-rater application and grouped recurring causes such as overlapping definitions, unclear boundaries, inconsistent residual categories, and different interpretations of the affected system aspect. These patterns informed the refinement reported in Section~\ref{sec:taxonomy:optimization}.

\section{Developing and Refining the Multidimensional Taxonomy}
\label{sec:taxonomy}

This section presents the results of the taxonomy development process. We first describe the criteria catalog derived from the literature analysis and bottom-up coding. We then show how the initial classification structure evolved into a multidimensional taxonomy, report its formative validation and refinement, and present the final taxonomy together with complementary post-refinement and application evidence.

\subsection{Criteria Catalog and Taxonomy Scope}
\label{sec:taxonomy:criteria}

The literature analysis and bottom-up coding produced a catalog of criteria for characterizing explanation needs. The catalog extends beyond the dimensions of the final taxonomy and includes information about the content of an explanation need, the review and application, linguistic properties, and the organizational context in which a need may be processed.

Table~\ref{tab:criteria-catalog} groups the identified criteria according to their analytical role. Content-level criteria formed the primary candidates for the taxonomy, while the remaining groups were retained as supplementary information for tasks such as detection, filtering, aggregation, or organizational decision making.

\begin{table*}[t]
\centering
\caption{Groups of criteria identified through the literature analysis and bottom-up analysis of app reviews. The taxonomy development focuses on content-level criteria that characterize an individual explanation need.}
\label{tab:criteria-catalog}
\begin{tabularx}{\textwidth}{p{0.22\textwidth} X X}
\toprule
\textbf{Criteria group} &
\textbf{Examples} &
\textbf{Role in this study} \\
\midrule

Content-level taxonomy candidates &
System components, external contexts, unexpected system behavior and bugs, software features, and temporal references &
Selected for taxonomy development because they characterize the semantic content of an individual explanation need \\

Explanation-need characterization &
Explicit or implicit expression, interaction type, and primary or secondary concern &
Retained as supplementary characterization or as part of the preceding identification of explanation needs \\

App and review metadata &
App name, genre, username, app version, device, operating system, and review date &
Excluded from the taxonomy because availability depends on the platform and dataset \\

Text-derived indicators and statistics &
Review length, punctuation, verbs, question words, sentiment, emotion, and user experience &
Retained as possible indicators for automated analysis but excluded because they do not directly describe the subject of the knowledge gap \\

Corpus-level information &
Frequency of an explanation need across several reviews &
Excluded because it requires aggregation across multiple reviews \\

External models and organizational factors &
ISO/IEC 25010 quality characteristics and estimated resolution costs &
Retained as possible mappings or decision-support information but excluded because they depend on external models or organizational estimates \\

\bottomrule
\end{tabularx}
\end{table*}

We selected criteria for the taxonomy when they described the semantic content or context of an individual explanation need, could be applied to a single need, supported categorical classification by human raters, and were sufficiently general across applications and platforms.

The distinction between content and metadata is particularly relevant for hardware. The device on which a review was submitted is metadata. Hardware can also be the subject of an explanation need, for example when a user asks about smartwatch compatibility. In this case, it characterizes the content of the need and can therefore be represented in the taxonomy.

Based on these conditions, subsequent development focused on content-level criteria describing temporal context, unexpected system behavior and bugs, software features, and affected system or external aspects.

\subsection{Initial Coding System}
\label{sec:taxonomy:initial}

During the first coding pass, each explanation need received a short code describing the aspect the user sought to understand. For example, reviews criticizing the relocation of a search bar could be represented by the code ``search bar moved.''

Similar codes were grouped into categories and then consolidated into higher-level categories. This process produced an initial coding system with 15 top-level categories and 38 categories in total, covering areas such as accounts, unexpected behavior and bugs, software features, content, security, commercial concerns, user interfaces, and updates.

The complete hierarchy appears in the replication package in \path{taxonomy/01-initial-coding-system.xlsx}.

The coding system revealed that several categories represented different classification perspectives. A need concerning a moved search bar, for example, can be both update-related and user-interface-related. A first restructuring made temporal context the top level of the hierarchy, but this merely made one perspective dominant while other relevant properties remained difficult to preserve.

\subsection{From a Single Hierarchy to Four Dimensions}
\label{sec:taxonomy:dimensions}

During the pre-study, the experts identified drawbacks of making temporal context the root of our interim hierarchy. We reorganized the coding system so that the same need could be retrieved through different properties.

Table~\ref{tab:traceability} traces how concepts from the reviewed classifications and groups of bottom-up codes contributed to these dimensions. Some source concepts contribute to more than one dimension, illustrating why the resulting perspectives are complementary rather than competing.

\begin{table*}[t]
\centering
\footnotesize
\setlength{\tabcolsep}{5pt}
\caption{Traceability from literature concepts and bottom-up code groups to the four dimensions of the taxonomy.}
\label{tab:traceability}
\begin{tabular}{@{}p{0.30\textwidth}p{0.22\textwidth}p{0.40\textwidth}@{}}
\toprule
\textbf{Source concept or code group} & \textbf{Origin} & \textbf{Dimension in the taxonomy} \\
\midrule
Update-related needs, references to upcoming versions & Bottom-up codes & Time Aspect \\
Unexpected behavior, incorrect output, inconsistent behavior, lost data & Bottom-up codes & Unexpected System Behavior \& Bugs \\
System behavior, errata & Literature~\cite{droste2024taxonomy,droste2025operationaltaxonomy,unterbusch2023explanation} & Unexpected System Behavior \& Bugs \\
Feature requests, changed and removed features, uncertainty about features & Bottom-up codes & Software Feature \\
Training, interaction, tutorial-related concerns & Literature~\cite{droste2024taxonomy,droste2025operationaltaxonomy,unterbusch2023explanation} & Software Feature, with the affected component in System Aspect \\
Account, login, notifications, synchronization, user interface, content, advertisements & Bottom-up codes & System Aspect, \emph{Internal} branch \\
Privacy and security, user interface, domain knowledge & Literature~\cite{droste2024taxonomy,droste2025operationaltaxonomy} & System Aspect, \emph{Internal} branch \\
Hardware compatibility, commercial concerns, customer support & Bottom-up codes & System Aspect, \emph{External} branch \\
Business concerns & Literature~\cite{unterbusch2023explanation} & System Aspect, \emph{External} branch \\
Design rationale, reasons for decisions & Literature~\cite{unterbusch2023explanation} and bottom-up codes & System Aspect, \emph{Development Rationale} \\
\midrule
Primary and secondary concern, dissatisfaction & Literature~\cite{unterbusch2023explanation} & Not represented as a dimension. Retained as supplementary characterization \\
App version, device, review date, sentiment, review length & Literature on app review mining & Not represented as a dimension. Retained as metadata and text-derived indicators \\
\bottomrule
\end{tabular}
\end{table*}

Every identified explanation need receives one label in each dimension. When the corresponding property is not expressed, a non-applicability label is assigned.

The multidimensional structure avoids forcing complementary properties into one hierarchy. A need concerning a search bar moved after an update can therefore preserve both its temporal context and its relation to the user interface.

\finding{Making time the root of our interim hierarchy gave it priority over other properties. The four-dimensional design records each selected property explicitly.}

\subsection{Formative Validation of the Initial Taxonomy}
\label{sec:taxonomy:validation}

The initial multidimensional taxonomy was formatively evaluated through the expert pre-study and an inter-rater application. Two raters independently classified 104 reviews and assigned one label in each dimension, resulting in 416 classification decisions per rater.

Table~\ref{tab:kappa} presents the number of agreements, percentage agreement, and Cohen's Kappa for each dimension based on the independent assignments before discussion of disagreements.

\begin{table*}[t]
\centering
\caption{Inter-rater agreement for the initial multidimensional taxonomy. Two raters independently classified 104 purposively selected reviews in four dimensions, resulting in 416 assignments per rater.}
\label{tab:kappa}
\small\setlength{\tabcolsep}{4pt}
\begin{tabular}{lrrrrl}
\toprule
\textbf{Dimension} &
\textbf{Agreements} &
\textbf{Disagreements} &
\textbf{Agreement} &
\textbf{Cohen's $\kappa$ [95\% CI]} &
\textbf{Interpretation~\cite{landis1977measurement}} \\
\midrule

Time Aspect &
101 &
3 &
97.12\% &
0.911 [0.791, 1.000] &
Almost perfect \\

Unexpected System Behavior \& Bugs &
81 &
23 &
77.88\% &
0.610 [0.459, 0.744] &
Substantial \\

Software Feature &
67 &
37 &
64.42\% &
0.467 [0.325, 0.599] &
Moderate \\

System Aspect &
65 &
39 &
62.50\% &
0.579 [0.470, 0.678] &
Moderate \\

\midrule

All assignments &
314 &
102 &
75.48\% &
\multicolumn{2}{c}{--} \\

\bottomrule
\end{tabular}
\end{table*}

Both raters agreed on all four labels for 35/104 reviews (33.65\%). They agreed on three dimensions for 41 reviews, two for 23, and one for five. The pooled result of 314/416 answers a different question from agreement on an entire profile.

Agreement was highest for Time Aspect and lower for Software Feature and System Aspect. The qualitative analysis showed that disagreements followed recurring patterns rather than isolated classification errors.

For Software Feature, disagreements mainly concerned whether a specific feature was addressed and how closely related forms of uncertainty should be distinguished. The initial categories \emph{Unclarity}, \emph{Missing}, and \emph{Definition} did not provide sufficiently clear boundaries.

For Unexpected System Behavior \& Bugs, all disagreements concerned whether a reported problem belonged to the dimension at all. The specific behavior categories were not confused with one another, indicating that the main ambiguity concerned the dimension boundary toward expected limitations and other concerns.

For System Aspect, disagreements frequently involved categories such as \emph{Content}, \emph{User Experience}, and \emph{Main App Functionality}. These cases indicated ambiguity about whether the category should represent the affected component, the user's evaluation, or the general purpose of the application.

In Software Feature, 27 of 37 disagreements paired \emph{Absent} with \emph{Unclarity}, and four paired \emph{Missing} with \emph{Unclarity}. In Behavior, 21 of 23 paired \emph{Absent} with \emph{Incorrect Result} and two with \emph{Performance Degradation}. In System Aspect, three pairs each accounted for five of 39 disagreements: Content/User Experience, Business/Security, and Business/Content. Time Aspect had three disagreements, all involving \emph{Absent}. Counts combine both directions. Rater 1 assigned \emph{Unclarity} to 69 reviews and rater 2 to 46, with 40 in common. Positive agreement for this label was $2(40)/(69+46)=69.6\%$.

The agreement results were used formatively to identify weaknesses in the initial structure and guide its refinement.

\finding{The initial taxonomy showed different levels of applicability across dimensions. In the purposively selected reviews, agreement ranged from 62.50\% to 97.12\%, and the disagreements revealed recurring problems in category boundaries rather than isolated classification errors.}

\subsection{Refinement of the Taxonomy}
\label{sec:taxonomy:optimization}

The refinement addressed ambiguities identified through expert feedback and the inter-rater application. Its main goals were to reduce overlap between categories, clarify dimension boundaries, and represent non-applicability consistently.

Table~\ref{tab:optimization} traces the main refinements from the initial structure to the final taxonomy and links each change to the corresponding evidence.

\begin{table*}[htp!]
\centering
\footnotesize
\caption{Evolution of the taxonomy: initial categories, observed problems, evidence from the formative validation (expert pre-study and inter-rater application with 104 reviews), and resulting final categories.}
\label{tab:optimization}
\begin{tabularx}{\textwidth}{p{0.20\textwidth} X p{0.22\textwidth} p{0.21\textwidth}}
\toprule
\textbf{Initial categories / structure} &
\textbf{Observed problem} &
\textbf{Evidence} &
\textbf{Final categories / structure} \\
\midrule

Residual labels \emph{Not Categorized}, \emph{No Feature Mentioned}, and \emph{Time Independent} (all dimensions) &
Non-applicability was represented through differently named categories with inconsistent meanings. &
Expert feedback, together with systematic rater asymmetry in the marginal distributions of the non-applicability labels in the behavior- and feature-related dimensions &
Consistent \emph{Present} and \emph{Absent} structure in the Time Aspect, Unexpected System Behavior \& Bugs, and Software Feature dimensions. The System Aspect dimension distinguishes \emph{Internal} system aspects from \emph{External} contexts \\

\emph{Future}, \emph{Update}, \emph{Time Independent} (Time Aspect) &
Temporal categories lacked precise definitions, although temporal references were generally recognizable. &
Only 3 disagreements (97.12\% agreement, $\kappa=0.911$), together with expert feedback &
\emph{Present} divided into \emph{Future Plans} and \emph{Update Issues}. \emph{Absent} for needs without a temporal reference \\

\emph{Performance Degradation}, \emph{Data Loss}, \emph{Incorrect Result}. The precursor codes \emph{Incorrect Output}, \emph{Unexpected Interaction Outcomes}, and \emph{Inconsistent Behavior} from the initial coding system had been consolidated as \emph{Incorrect Result} (Unexpected System Behavior \& Bugs) &
The boundary of the dimension toward expected limitations and other concerns was unclear. The raters disagreed on whether a described problem constituted unexpected behavior at all. &
23 disagreements (77.88\% agreement, $\kappa=0.610$). All disagreements opposed a behavior category to the non-applicability label, while the behavior categories were never confused with one another &
Sharpened dimension definition with an explicit boundary toward expected limitations. \emph{Present} retains the categories \emph{Performance Degradation}, \emph{Data Loss}, and \emph{Incorrect Result} \\

\emph{Unclarity}, \emph{Missing}, and \emph{Definition}, together with separate categories for changed and removed features (Software Feature) &
The three categories described closely related forms of uncertainty without clear boundaries. The boundary between \emph{Unclarity} and reviews without a feature reference was applied inconsistently. &
37 disagreements (64.42\% agreement, $\kappa=0.467$). Most disagreements concerned whether a specific feature was addressed at all, supported by qualitative examples of users unable to find a function &
Uncertainty-related categories consolidated as \emph{Feature Confusion}. Changed and removed features combined as \emph{Feature Modification}. \emph{Absent} for needs without a feature reference \\

\emph{Content}, \emph{User Experience}, \emph{Main App Functionality}, and further mixed categories (System Aspect) &
Categories mixed the affected component, the user's evaluation, and the general purpose of the application. Some subcategories added complexity without improving the classification. &
39 disagreements (62.50\% agreement, $\kappa=0.579$), together with a qualitative analysis of boundary cases between content, interface, and application function &
Subject-focused definitions and a simplified category set. \emph{Hardware}, \emph{Business}, and \emph{Customer Support} retained as \emph{External} contexts \\

\bottomrule
\end{tabularx}
\end{table*}

The refinement introduced a consistent \emph{Present} and \emph{Absent} structure where properties may not be expressed. System Aspect instead distinguishes \emph{Internal} system aspects from \emph{External} contexts because every explanation need refers to some subject.

The final refinement also reduced overlapping and residual categories and clarified definitions. Annotation guidelines were then developed with category definitions, assignment rules, and representative examples.

\finding{The formative evaluation directly informed the refinement of the taxonomy. Overlapping categories were consolidated, dimension boundaries were clarified, non-applicability was represented consistently, and internal system aspects were separated from external contexts.}
\subsection{Final Taxonomy and Annotation Guidelines}
\label{sec:taxonomy:final}

\begin{figure*}[pos=!tbp]
    \centering
    \includegraphics[width=\textwidth]{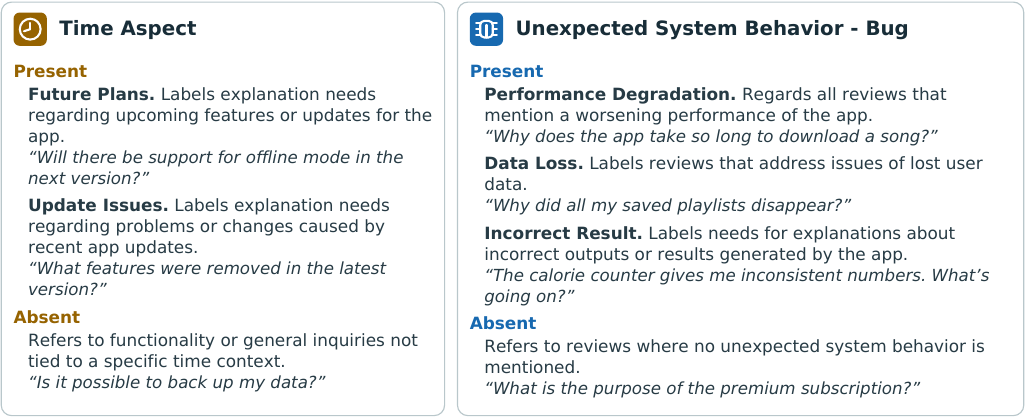}
    \caption{Annotation guidelines for the \emph{Time Aspect} and \emph{Unexpected System Behavior \& Bugs} dimensions.}
    \label{fig:guideline-time-behavior}
\end{figure*}

\begin{figure}[pos=!tbp]
    \centering
    \includegraphics[width=\columnwidth]{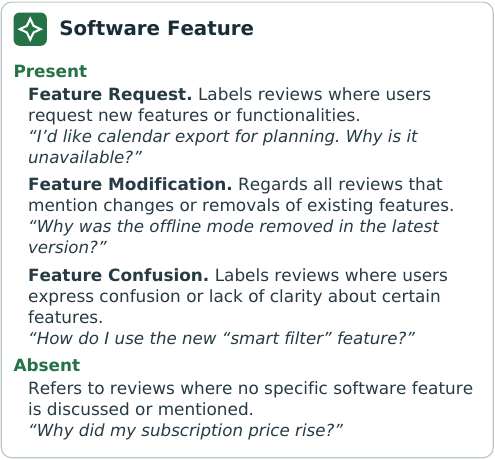}
    \caption{Annotation guideline for the \emph{Software Feature} dimension. Apply these labels to an identified explanation need. For \emph{Feature Request}, the review must also indicate what the user seeks to understand about the requested functionality.}
    \label{fig:guideline-feature}
\end{figure}

\begin{figure*}[pos=!tbp]
    \centering
    \includegraphics[width=\textwidth]{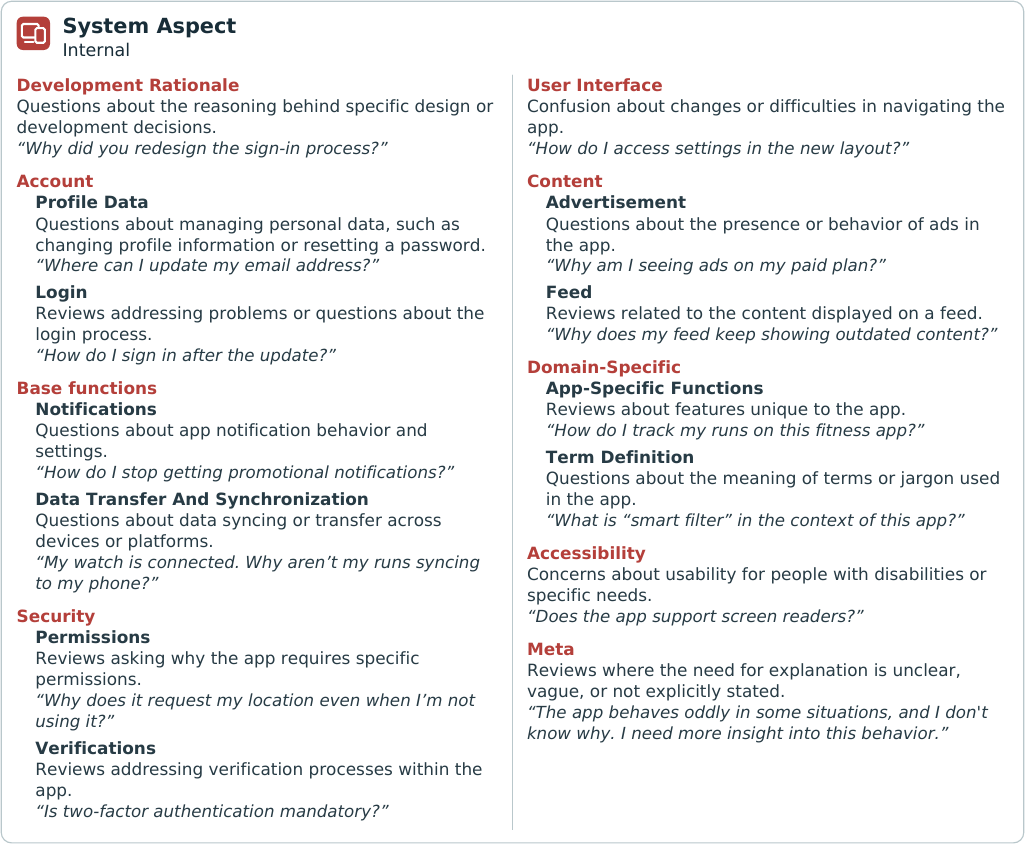}
    \caption{Annotation guideline for the \emph{Internal} branch of the \emph{System Aspect} dimension.}
    \label{fig:guideline-system-internal}
\end{figure*}

\begin{figure*}[pos=!tbp]
    \centering
    \includegraphics[width=\textwidth]{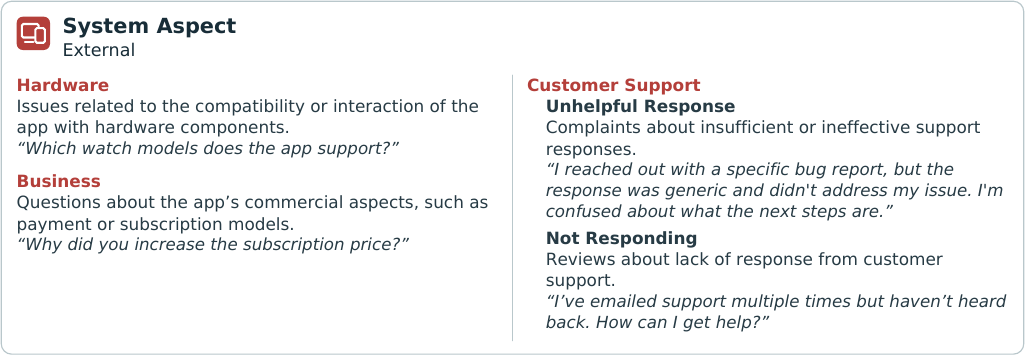}
    \caption{Annotation guideline for the \emph{External} branch of the \emph{System Aspect} dimension.}
    \label{fig:guideline-system-external}
\end{figure*}

The final taxonomy consists of four complementary dimensions: \emph{Time Aspect}, \emph{Unexpected System Behavior \& Bugs}, \emph{Software Feature}, and \emph{System Aspect}. Its application presupposes that an explicit or implicit explanation need has already been identified. Figures~\ref{fig:guideline-time-behavior} to~\ref{fig:guideline-system-external} present the complete category structure, definitions, and examples.

In the first three dimensions, \emph{Absent} means that the property is absent from an already identified need. First identify the words that reveal what the user seeks to understand. A desired change, complaint, or question phrased as a request for action is insufficient on its own. For an implicit need, record the words that support the inferred knowledge gap. For instance, \enquote{Please move Settings back} expresses a preference. \enquote{I cannot tell where Settings went after the redesign} identifies missing knowledge. The revised examples and tie rules below clarify the guidance. The agreement values refer to the guideline versions used in the respective studies.

\paragraph{Time Aspect.}

The Time Aspect dimension captures release- or development-related temporal references. It distinguishes \emph{Future Plans}, \emph{Update Issues}, and \emph{Absent}. General temporal expressions such as a problem occurring yesterday do not qualify unless they refer to an update or planned change.

\paragraph{Unexpected System Behavior \& Bugs.}

This dimension captures behavior that differs from the user's expectations. It distinguishes \emph{Performance Degradation}, \emph{Data Loss}, \emph{Incorrect Result}, and \emph{Absent}. The dimension concerns unexpected system behavior rather than expected limitations or general dissatisfaction.

\paragraph{Software Feature.}

The Software Feature dimension captures whether and how an explanation need refers to a specific software feature. It distinguishes \emph{Feature Request}, \emph{Feature Modification}, \emph{Feature Confusion}, and \emph{Absent}.

\paragraph{System Aspect.}

The System Aspect dimension identifies the subject of the explanation need. Unlike the other dimensions, it does not use an \emph{Absent} label because every explanation need refers to a subject. Its first level distinguishes \emph{Internal} system aspects from \emph{External} contexts. The complete hierarchy is shown in Figures~\ref{fig:guideline-system-internal} and~\ref{fig:guideline-system-external}.

The external branch preserves needs concerning hardware, business decisions, or customer support rather than assigning them to an undifferentiated residual category. This distinction also supports routing concerns to organizational units outside the development team.

\paragraph{Labeling Rules.}

For hierarchical System Aspect categories, annotators assign a leaf category. Parent categories are used only for aggregation. When a need refers to several concepts within one dimension, the category corresponding to the subject of the explanation need is assigned rather than a concept mentioned only in passing. \emph{Meta} is used only when the affected system aspect cannot be determined from the review.

For a review with multiple possible needs, write the question that an answer would resolve. Repeated descriptions of the same gap stay together. Split when answering one question leaves another distinct question unanswered. Retain the review ID and text span for each need, then assign each its own four-label profile. One archived review asks what happened to the music and, later, why there is no comet. These are two needs in one review. In a retrospective search of the first-iteration MAXQDA code tree, excluding its no-need branch, 168 of 2{,}133 coded documents had at least two non-overlapping coded passages. Separate passages can elaborate the same gap and one passage can contain two questions. These markings show how different passages could be retained during development. The full-corpus frequency comes from the per-review need annotations in Section~\ref{sec:studydesign:dataset}.

One label per dimension keeps grouping counts interpretable at the level of the need. For a question about a changed login, code \emph{Development Rationale} if the user wants the reason for the design decision, or \emph{Login} if the user needs to understand how to sign in. For a smartwatch, distinguish device compatibility (\emph{Hardware}) from how data travel between supported devices (\emph{Data Transfer and Synchronization}). When two labels remain equally justified within one need, record the alternative and rationale for adjudication, preserve both raters' initial judgments, and flag unresolved cases.

\paragraph{Customization and Extension.}

The four dimensions and their core categories form the normative taxonomy. Application-specific profiles may add subcategories below existing core categories provided that each added category maps to exactly one core category. Core categories must not be removed, renamed, or moved between dimensions. This preserves comparability while allowing application-specific granularity. Such an extension was used in the industrial study, where the application-specific category \emph{Report Center} was introduced~\cite{obaidi2025elicitation}.

\paragraph{Compatibility with Existing Taxonomies.}

Categories from the taxonomies by Droste et al.~\cite{droste2024taxonomy,droste2025operationaltaxonomy} and Unterbusch et al.~\cite{unterbusch2023explanation} can be mapped to corresponding categories of the multidimensional taxonomy. The remaining dimensions can then be annotated from the original explanation need to add complementary information. Complete mapping tables are provided in the replication package.

The formative sheet retains the prior subject alongside both four-dimensional profiles. Among 23 rows marked simply \emph{Business} in the prior field, both raters assigned \emph{Account} as the System Aspect in two, \emph{Content} in three, and customer support in two. An analyst could route the two account cases separately from the two customer-support cases. Across 104 reviews, rater 1 marked at least two of the three optional properties as present in 57 cases, and rater 2 did so in 43. We excluded the obligatory System Aspect. The prior field was used for sampling and sometimes contains combined labels. These results illustrate co-occurrence and grouping. They do not measure information loss under an independently applied baseline or routing accuracy.
\subsection{Agreement in the Multidimensional Gold-Standard Dataset}
\label{sec:taxonomy:postrefinement-results}

The final taxonomy was subsequently applied as the labeling scheme of the separate multidimensional gold-standard dataset described in Section~\ref{sec:studydesign:postrefinement}~\cite{obaidi2026multigolddataset}. We use the archived rating decisions to estimate category-level agreement and its uncertainty for this separate application of the final taxonomy.

Table~\ref{tab:postrefinement-agreement} reports binary branch-level results. We recomputed the two round estimates before averaging, avoiding intermediate rounding. The analysis covers 1{,}571 aligned need passages from 1{,}410 reviews in round 1 and 1{,}614 passages from 1{,}433 other reviews in round 2, each with three votes. These are the units entering taxonomy agreement within the 5{,}004-review detection corpus. Among the 2{,}843 included reviews, 291 had more than one aligned passage. This describes the separate campaign, with its own identification and alignment procedures.

\begin{table}[pos=!htbp]
\centering\small
\caption{Binary branch agreement in the separate dataset. Each estimate is the unweighted mean of the two main annotation rounds. Intervals resample whole reviews within rounds.}
\label{tab:postrefinement-agreement}
\begin{tabular}{@{}lrr@{}}
\toprule
Dimension & Fleiss' $\kappa$ & 95\% CI \\
\midrule
Time Aspect & 0.653 & [0.621, 0.683] \\
\makecell[l]{Unexpected System\\Behavior \& Bugs} & 0.724 & [0.703, 0.745] \\
Software Feature & 0.519 & [0.496, 0.542] \\
System Aspect & 0.779 & [0.755, 0.802] \\
\bottomrule
\end{tabular}
\end{table}

Table~\ref{tab:category-agreement} gives all category results, with the number of passages receiving at least one positive vote. Each leaf is assessed as one versus the rest, and each parent aggregates its subtree. For \emph{Meta}, 81 passages received at least one positive vote, with 102 positive votes in total. Mean Fleiss' $\kappa$ was 0.254 [0.152, 0.337], AC1 was 0.98345, and pairwise agreement was 98.381\%. For \emph{Feature Confusion}, 1{,}351 passages received a positive vote and $\kappa$ was 0.406 [0.378, 0.432]. This difficulty persists in a frequently used category. Every reported category had positive votes, which shows observed use within this corpus. It does not establish coverage of needs excluded at the identification stage or outside the sampled applications.
\begin{table*}[pos=!tp]\centering\small
\caption{Category-level inter-rater agreement in the multidimensional gold-standard dataset. $n_+$ is the number of passages with at least one positive vote across both rounds. The denominator for every binary decision is 3{,}185 aligned passages from 2{,}843 reviews. AC1 and agreement are means of round estimates.}
\label{tab:category-agreement}
\begin{tabular*}{\textwidth}{@{\extracolsep{\fill}}lrrrrr@{}}\toprule
Category or branch & $n_+$ & Fleiss$^\prime$ $\kappa$ & 95\% CI & AC1 & Agreement (\%) \\ \midrule
\multicolumn{6}{@{}l}{\textbf{Time Aspect}} \\
Present vs. Absent & 596 & 0.653 & [0.621, 0.683] & 0.896 & 92.02 \\
Update Issues & 455 & 0.671 & [0.634, 0.704] & 0.926 & 93.98 \\
Future Plans & 161 & 0.541 & [0.480, 0.601] & 0.971 & 97.29 \\
\addlinespace[2pt]
\multicolumn{6}{@{}l}{\textbf{Unexpected System Behavior \& Bugs}} \\
Present vs. Absent & 1287 & 0.724 & [0.703, 0.745] & 0.790 & 88.08 \\
Performance Degradation & 671 & 0.513 & [0.474, 0.548] & 0.869 & 89.72 \\
Incorrect Result & 864 & 0.491 & [0.462, 0.518] & 0.797 & 85.56 \\
Data Loss & 101 & 0.711 & [0.639, 0.772] & 0.987 & 98.74 \\
\addlinespace[2pt]
\multicolumn{6}{@{}l}{\textbf{Software Feature}} \\
Present vs. Absent & 2074 & 0.519 & [0.496, 0.542] & 0.524 & 76.08 \\
Feature Request & 615 & 0.714 & [0.685, 0.741] & 0.905 & 92.88 \\
Feature Modification & 396 & 0.583 & [0.542, 0.621] & 0.928 & 93.89 \\
Feature Confusion & 1351 & 0.406 & [0.378, 0.432] & 0.645 & 77.81 \\
\addlinespace[2pt]
\multicolumn{6}{@{}l}{\textbf{System Aspect}} \\
Internal vs. External & 2767 & 0.779 & [0.755, 0.802] & 0.905 & 93.37 \\
Development Rationale & 468 & 0.476 & [0.440, 0.511] & 0.899 & 91.58 \\
Account & 207 & 0.753 & [0.707, 0.795] & 0.974 & 97.69 \\
Profile Data & 112 & 0.573 & [0.484, 0.646] & 0.982 & 98.24 \\
Login & 112 & 0.784 & [0.727, 0.835] & 0.987 & 98.80 \\
Base Functions & 484 & 0.607 & [0.576, 0.637] & 0.908 & 92.52 \\
Notifications & 79 & 0.667 & [0.589, 0.738] & 0.988 & 98.87 \\
Data Transfer and Synchronization & 408 & 0.602 & [0.569, 0.633] & 0.923 & 93.52 \\
Security & 140 & 0.776 & [0.726, 0.822] & 0.984 & 98.51 \\
Permissions & 92 & 0.801 & [0.739, 0.852] & 0.991 & 99.10 \\
Verifications & 52 & 0.677 & [0.582, 0.760] & 0.992 & 99.24 \\
User Interface & 896 & 0.582 & [0.552, 0.611] & 0.828 & 87.80 \\
Content & 1195 & 0.601 & [0.575, 0.625] & 0.755 & 84.91 \\
Advertisement & 148 & 0.832 & [0.785, 0.872] & 0.987 & 98.81 \\
Feed & 1065 & 0.566 & [0.539, 0.593] & 0.776 & 85.36 \\
Domain-Specific & 516 & 0.377 & [0.335, 0.419] & 0.887 & 90.47 \\
App-Specific Functions & 423 & 0.318 & [0.274, 0.362] & 0.907 & 91.88 \\
Term Definition & 103 & 0.524 & [0.449, 0.589] & 0.981 & 98.17 \\
Accessibility & 41 & 0.622 & [0.472, 0.731] & 0.994 & 99.37 \\
Meta & 81 & 0.254 & [0.152, 0.337] & 0.983 & 98.38 \\
Hardware & 108 & 0.583 & [0.483, 0.661] & 0.982 & 98.30 \\
Business & 535 & 0.806 & [0.780, 0.830] & 0.941 & 95.45 \\
Customer Support & 117 & 0.728 & [0.665, 0.781] & 0.985 & 98.57 \\
Unhelpful Response & 81 & 0.686 & [0.602, 0.753] & 0.988 & 98.89 \\
Not Responding & 41 & 0.696 & [0.591, 0.782] & 0.995 & 99.47 \\
\bottomrule\end{tabular*}
\end{table*}
\subsection{Industrial Application of the Refined Taxonomy}
\label{sec:taxonomy:application}

Obaidi et al. used a two-part scheme to code explanation needs elicited for a personnel management system in an industrial case study~\cite{obaidi2025elicitation,obaidi2026eliciting}. The researchers used Droste's taxonomy during elicitation. For coding and identifying distinct needs, they added fields for time, software feature, and system aspect, together with software-specific categories. The companion dataset describes a combined string of coding fields used to identify matching needs~\cite{industrialartifact2025}. This was an application-specific implementation.

One engineer classified the statements and consulted a second engineer in uncertain cases. Using the application-specific extension, the industrial study grouped 471 survey statements into 364 distinct explanation needs~\cite{obaidi2025elicitation}. These counts describe the output of grouping. The journal account compares two focus groups, 18 interviews, and a survey with 188 valid responses. Its agreement analysis concerns how participants and analysts assigned elicited needs to categories. The study materials document the elicitation and coding procedures~\cite{obaidi2026eliciting,obaidi2026replication}. The grouping task had no alternative coding scheme, independent reference grouping, or timing data per coding decision. We therefore cannot estimate deduplication accuracy, effort saved, or improvements in elicited requirements from these counts.

\section{Discussion}
\label{sec:discussion}

This section answers the research questions, discusses the implications of the multidimensional taxonomy, outlines directions for future work, and examines threats to validity.

\subsection{Answering the Research Questions}
\label{sec:discussion:rqs}

\paragraph{RQ1: Which criteria can be derived from existing classifications and app reviews for characterizing explanation needs?}

The analysis produced a broad criteria catalog covering semantic properties of explanation needs as well as metadata, linguistic indicators, corpus-level information, and organizational factors. For the taxonomy itself, we retained content-level criteria that characterize an individual explanation need independently of platform-specific or organizational information. These criteria concern temporal context, unexpected system behavior and bugs, software features, and the affected system or external aspect.

\paragraph{RQ2: How can suitable content-level criteria be organized into a multidimensional taxonomy of explanation needs?}

The selected criteria represent complementary properties that may occur simultaneously and can be useful as separate retrieval fields. We organized them into four dimensions: \emph{Time Aspect}, \emph{Unexpected System Behavior \& Bugs}, \emph{Software Feature}, and \emph{System Aspect}. Each explanation need receives one label in every dimension, preserving complementary information for subsequent software engineering analysis.

\paragraph{RQ3: Which ambiguities emerge when the initial taxonomy is reviewed and independently applied, and how can the taxonomy be refined accordingly?}

The formative evaluation primarily revealed ambiguities in category and dimension boundaries. Software Feature showed uncertainty about whether a feature was addressed and how related forms of feature uncertainty should be distinguished. System Aspect contained overlapping interpretations of the affected subject, while disagreements in Unexpected System Behavior \& Bugs mainly concerned the boundary of the dimension itself. Time Aspect was comparatively stable.

These findings led to the consolidation of feature-related categories, clearer dimension boundaries, consistent representation of non-applicability, restructuring of System Aspect into internal and external branches, and clarified annotation guidelines.

\FloatBarrier
\subsection{Implications of the Multidimensional Taxonomy}
\label{sec:discussion:implications}

The taxonomy should be understood as an intermediate representation for processing post-release user feedback. Existing explanation-need taxonomies describe general subjects or concerns~\cite{droste2024taxonomy,droste2025operationaltaxonomy,unterbusch2023explanation}. The multidimensional structure complements these approaches by preserving several properties of the same need for subsequent analysis.

This distinction can support different software engineering responses. Feature requests may inform requirements and product planning, feature confusion may indicate documentation or interface problems, and unexpected behavior may trigger technical investigation. Security, business, or customer-support concerns may require other stakeholders. The taxonomy does not prescribe these decisions but preserves information that can support them. Guideline-based approaches can subsequently assist in formulating explainability requirements and explanations~\cite{obaidi2026guidelineReq,obaidi2026guidelineExp}.

The taxonomy deliberately separates semantic content from supplementary information such as metadata, sentiment, linguistic indicators, and frequency. This keeps the core structure independent of platform-specific and organization-specific information. For sentiment in particular, Herrmann et al.~\cite{herrmann2022subjectivity} found that readers of software-project messages often judged their tone differently from the polarity assigned in the original datasets. At the same time, application-specific subcategories can extend the taxonomy while retaining the four-dimensional core, as demonstrated in the industrial study~\cite{obaidi2025elicitation}.

For research, the finite category sets and annotation guidelines provide an operational target for automated classification. The public multidimensional dataset labeled with the taxonomy can support the training and evaluation of classifiers for detecting and characterizing explanation needs~\cite{obaidi2026multigolddataset}.

\subsection{Implications of the Agreement Evidence}
\label{sec:discussion:agreement}

The two agreement analyses have different empirical roles. The formative 104-review application concerns the initial taxonomy and directly informed its refinement. The reused category-level agreement evidence from the separate multidimensional dataset instead concerns the final taxonomy and characterizes its application in an independent annotation campaign.

The formative results indicate that some dimensions require more interpretation than others. Temporal references were comparatively easy to recognize, while feature-related and system-related classifications produced more boundary ambiguities. The qualitative disagreement analysis was therefore particularly important because it identified overlapping categories, unclear dimension boundaries, and inconsistent handling of non-applicability. These observations motivated the refinement described in Section~\ref{sec:taxonomy:optimization}.

The separate multidimensional dataset provides complementary post-refinement evidence for the final taxonomy~\cite{obaidi2026multigolddataset}. Agreement was moderate to substantial for most categories, although some System Aspect categories and \emph{Feature Confusion} remained more difficult to distinguish consistently. Rare categories also demonstrate that Kappa should be interpreted together with percentage agreement and Gwet's AC1. Thus, the formative evidence explains why the initial taxonomy was refined, while the reused evidence indicates how consistently the final structure was subsequently applied.

\subsection{Future Work}
\label{sec:discussion:futurework}

Future evaluations should focus particularly on categories that showed lower agreement in the complementary post-refinement evidence. Studies involving other raters, application domains, platforms, and languages could determine whether these difficulties arise primarily from category definitions, prevalence, or corpus characteristics.

Future work should test segmentation on unfiltered reviews and report agreement on detecting needs and identifying their spans before measuring agreement on the taxonomy labels. Controlled settings may provide complementary evidence on mental-model conflicts~\cite{droste2026misunderstandings}.

The finite category sets and annotation guidelines also provide a basis for automated classification. The multidimensional dataset offers a corresponding training and evaluation resource~\cite{obaidi2026multigolddataset}. Future studies could evaluate machine learning and large language model approaches for detecting explanation needs and assigning labels in each dimension, and connect these classifications to the generation of explainability requirements and explanations~\cite{obaidi2025explainreqs}.

Finally, supplementary criteria such as metadata, sentiment, linguistic indicators, frequency, and organizational information could be combined with the taxonomy in downstream applications and evaluated independently.

\subsection{Threats to Validity}
\label{sec:discussion:threats}

We discuss threats to validity according to the categories proposed by Wohlin et al.~\cite{wohlin2012experimentation}.

\subsubsection{Construct Validity}

A first threat concerns the definition of an explanation need. Explicit questions are often easy to identify, but implicit needs require interpretation. Reviews that express dissatisfaction or confusion may not always contain a genuine request for explanatory information. We reduced this threat by using an existing gold-standard dataset labeled for explanation needs~\cite{goldstandard-explain-zenodo2024}. However, the taxonomy depends on the correctness of these initial labels.

A second threat concerns the selection of taxonomy dimensions. The criteria catalog also contained metadata, linguistic indicators, quality-model mappings, and other information. We selected dimensions based on their suitability for characterizing individual explanation needs, but other researchers may define the scope differently.

A third threat concerns the interpretation of \emph{Absent}. In the final taxonomy, \emph{Absent} is used in the Time Aspect, Unexpected System Behavior \& Bugs, and Software Feature dimensions and means that the property represented by the dimension is not expressed in the explanation need. It does not mean that the review lacks an explanation need. The System Aspect dimension avoids this ambiguity by distinguishing internal system aspects from external contexts instead of using an \emph{Absent} label. These distinctions must be communicated clearly to annotators.

\subsubsection{Internal Validity}

The bottom-up coding was conducted collaboratively by two researchers. This reduces dependence on the interpretation of a single coder, because codes, category boundaries, and unclear cases were discussed during the coding process. However, the complete corpus was not independently double-coded by both researchers before consensus. Shared assumptions may therefore have influenced the resulting coding system. The expert pre-study and the later independent inter-rater application provided additional perspectives, but they cannot eliminate all effects of decisions made during the collaborative bottom-up analysis.

The inter-rater validation used a purposefully selected subset of 104 reviews. The selection aimed to cover the categories of the initial coding system and the taxonomy by Droste et al.~\cite{droste2024taxonomy,droste2025operationaltaxonomy}. This strategy supported category coverage, but the subset may not represent the natural distribution of categories in the full dataset.

The raters may also have shared background knowledge due to their research context. Raters from other institutions or professional backgrounds might interpret the categories differently. The three raters underlying the reused evidence were not involved in taxonomy development, and the separate dataset contains no applications from the development dataset. However, they received training and worked with the final guidelines. The reused agreement evidence therefore characterizes the application of a fixed scheme after training rather than its use without prior introduction.

\subsubsection{Conclusion Validity}

The formative agreement results are based on 104 reviews and two raters. This sample was designed to identify recurring ambiguities in the initial taxonomy rather than to estimate the reliability of the refined category structure.

Cohen's Kappa depends on category prevalence and marginal distributions. A lower Kappa does not necessarily indicate that a dimension is unusable. Percentage agreement, category frequencies, and the qualitative causes of disagreement must be considered together.

The marginal distributions also indicate a systematic difference between the two raters. One rater assigned the label for non-applicability considerably more often than the other in the behavior-related and feature-related dimensions, which affects both the observed agreement and Cohen's Kappa.

Because the formative results concern the initial taxonomy, they cannot be transferred directly to the refined structure. We therefore consider reused category-level agreement evidence from the separate 5{,}004-review dataset~\cite{obaidi2026multigolddataset}. This provides a broader indication of consistent application of the final taxonomy, although agreement still varies across categories and does not eliminate concerns about annotation reliability.

The separate multidimensional dataset was purposefully constructed as a gold standard rather than as a naturally distributed random sample of all app-store reviews. Rare or broad categories may consequently still require further evaluation in other datasets and application domains. The reanalysis reports Fleiss' Kappa, Gwet's AC1, and pairwise agreement per category, with review-cluster bootstrap intervals for the round-mean Kappa values. These intervals are conditional on the aligned needs and raters in this dataset. Small differences between individual categories should therefore not be overinterpreted.

\subsubsection{External Validity}

The taxonomy was developed from a gold-standard dataset of English-language app reviews~\cite{goldstandard-explain-zenodo2024}, and its subsequent applications also concern English-language reviews and a German industrial context. Although several applications are represented, the findings may not generalize to all domains. Enterprise software, safety-critical systems, public-sector applications, or specialized professional tools may contain additional explanation needs. Research on video games, for example, reports domain-specific needs~\cite{droste2026immersive}.

The app-review evidence specifically concerns naturally occurring English-language feedback. The German industrial survey elicited statements in a different context and used an application-specific extension.

The taxonomy may also require adaptation for other feedback channels. Support tickets, forum posts, social media messages, and interview transcripts differ from app reviews in length, context, and available metadata. Explanations in developer forums, for instance, are shaped by factors such as length, code inclusion, and author reputation~\cite{obaidi2026usefulness}.

Finally, the taxonomy focuses on explanation needs expressed after users have interacted with software. It does not directly address explanation needs anticipated during requirements elicitation or explanation needs detected through runtime behavior, for which catalogs of user-based indicators have been proposed~\cite{deters2025explanationcatalog}.

\section{Conclusion}
\label{sec:conclusion}

Software engineering teams receive large volumes of post-release user feedback, but its usefulness depends on whether the expressed concerns can be represented in a form that supports analysis and action. Explanation needs are particularly informative because they reveal what users seek to understand about software behavior, features, changes, interfaces, security decisions, and surrounding contexts. An analyst may need to retrieve the same concern by more than one of these properties.

We developed and refined a multidimensional taxonomy that represents each identified explanation need along four complementary dimensions: \emph{Time Aspect}, \emph{Unexpected System Behavior \& Bugs}, \emph{Software Feature}, and \emph{System Aspect}. The taxonomy combines concepts from existing classifications with joint bottom-up coding of development reviews. The consolidated source corpus records 3{,}298 needs in 2{,}185 of 4{,}495 reviews. An expert pre-study and an independent inter-rater application to 104 purposively selected reviews provided formative evidence for the initial taxonomy. Agreement ranged from 62.50\% to 97.12\% across dimensions (Cohen's $\kappa=0.467$--$0.911$), and the disagreement analysis directly informed the consolidation of categories, clarification of dimension boundaries, consistent handling of non-applicability, restructuring of System Aspect, and development of annotation guidelines.

An application-specific extension was used in an industrial elicitation study, and the refined taxonomy supplied the labeling scheme of a separate multidimensional app-review dataset~\cite{obaidi2025elicitation,obaidi2026multigolddataset}. The industrial study provides complementary application-based evidence. Our reanalysis of category-level agreement in the separate 5{,}004-review annotation campaign provides complementary evidence on the final taxonomy, with Fleiss' $\kappa$ ranging from 0.41 to 0.71 across the first three dimensions and from 0.25 to 0.83 for System Aspect~\cite{obaidi2026multigolddataset}. The construction and broader analysis of that dataset remain separate research outputs and are not contributions of the present article.

Together, the taxonomy-development process, formative evaluation, complementary post-refinement agreement evidence, and industrial application support the final multidimensional artifact from different perspectives. The remaining variation in category-level agreement also identifies categories whose boundaries should be examined further in future applications and datasets.

A matched comparison on fresh material is needed to quantify the change in reliability after refinement. A comparative engineering task with an external reference or an effort measure would be needed to establish a downstream benefit.

\section*{Data Availability Statement}
The replication package contains the coding schemes, annotation guidelines, formative agreement data, mappings to prior taxonomies, and analysis scripts. Access to the package is currently limited to peer review. The separate multidimensional gold-standard dataset is publicly available~\cite{obaidi2026multigolddataset}.

\section*{Declaration of Competing Interest}
The authors declare that they have no known competing financial interests or personal relationships that could have appeared to influence the work reported in this paper.

\section*{Acknowledgment}
This work was funded by the Deutsche Forschungsgemeinschaft (DFG, German Research Foundation) under Grant No. 470146331, project softXplainer (2026--2028).

\section*{CRediT Authorship Contribution Statement}
\textbf{Martin Obaidi:} Conceptualization, Data curation, Formal analysis, Investigation, Methodology, Resources, Software, Validation, Visualization, Writing -- original draft, Writing -- review \& editing.

\textbf{Marc Herrmann:} Writing -- review \& editing.

\textbf{Kurt Schneider:} Supervision, Funding acquisition.




\end{document}